\documentclass{article}
\usepackage{graphicx}
\usepackage{dcolumn}
\usepackage{bm}
\usepackage{hyperref}
\usepackage{epstopdf}
\usepackage{amsmath, amsthm}
\usepackage{xcolor}
\usepackage[superscript]{cite}
\usepackage{authblk}

\usepackage[caption=false]{subfig}
\def \bea {\begin{eqnarray}}
\def \eea {\end{eqnarray}}
\def \be {\begin{equation}}
\def \ee {\end{equation}}
\def \asi{{\it a}-Si}
\def \gese3ag{GeSe$_3$Ag}
\def \aggese3{Ag$_x$(GeSe$_3$)$_{1-x}$}
\def \abfear{{\it ab initio} FEAR}
\def \asio2{{\it a}-SiO$_2$} 

\def \deg {${^\circ}$}

\title{Inversion of diffraction data for amorphous materials}
\author{Anup Pandey}
\affil{Department of Physics and Astronomy,  Ohio University, Athens OH 45701, USA}
\author{Parthapratim Biswas}
\affil{Department of Physics and Astronomy, The University of Southern Mississippi, Hattiesburg MS 39406, USA}
\author{D. A. Drabold}
\affil{Department of Physics and Astronomy, Ohio University, Athens OH 45701, USA}

\date{\today}
\begin{document}
\maketitle
\begin{abstract}
The general and practical inversion of diffraction data -- producing a computer model
correctly representing the material explored --  is an important unsolved problem for 
disordered materials.
Such modeling should proceed by using our full knowledge base, 
both from experiment and theory. In this paper, we describe a robust 
method to jointly exploit the power of {\it ab initio} atomistic 
simulation along with the information carried by diffraction data. 
The method is applied to two very different systems:  amorphous silicon 
and two compositions of a solid electrolyte memory material silver-doped GeSe$_3$. 
The technique is easy to implement, is faster and yields results 
much improved over conventional simulation methods for the materials 
explored. By direct calculation, we show that the method works for both poor and
excellent glass forming materials. It offers a means to add {\it a priori} 
information in first principles modeling of materials, and represents a significant 
step toward the computational design of non-crystalline materials 
using accurate interatomic interactions and experimental information. 
\end{abstract}


On the eve of the First World War, William Lawrence Bragg and his father, 
William Henry Bragg, exposed crystalline solids to X-rays and discovered 
what we now call ``Bragg diffraction", strong reflection at particular incident 
angles and wavelengths. These ``Bragg peaks" were sharply defined and, 
when analyzed with a wave theory of the X-rays, led to clear evidence of order 
in the crystalline state\cite{bragg}. By analyzing the diffraction 
angles at which the peaks appeared and the wavelength of the X-rays, 
the full structure of the crystal 
could be ascertained. In the language of modern solid state physics, 
the X-ray structure factor of a single crystal consists of a 
sequence of sharp spikes, which are broadened in a minor way by 
thermal effects. The information obtained from this 
palisade of delta functions, arising from a crystal, is sufficient for 
the determination of atomic structure of crystal uniquely. 
The rapid development of X-ray Crystallography in the past several decades 
had made it possible to successfully determine the structure of complex 
protein molecules, with more than $10^5$ atoms, leading to the formation 
of a new branch of protein crystallography in structural biology\cite{kendrew}.  

In contrast with crystals, amorphous materials and liquids have structure 
factors that are smooth, and thus contain far less specific 
information about structure. The lack of sharp peaks principally 
originates from the presence of local atomic ordering 
in varying length scales, and no long-range order in the amorphous state.  The resulting structure factor 
is one-dimensional and is effectively a sum rule that must be satisfied by the three-dimensional 
amorphous solids.  This presents a far more difficult problem of structural determination 
of amorphous solids that requires the development of new tools and reasoning 
to obtain realistic structural models. A natural approach to 
address the problem is to carry out computer simulations, either 
employing molecular dynamics or Monte Carlo, with suitable 
interatomic potentials. We have called this approach the ``simulation paradigm"\cite{dad_eur} 
elsewhere. By contrast, the other limit is to attempt to invert 
the diffraction data by ``Reverse Monte Carlo" (RMC) or otherwise 
without using any interatomic potential but information only~\cite{mcgreevy, biswas_rmc}. 
This we have called the ``information paradigm"\cite{dad_eur}. The information paradigm 
in its purest form produces models reproducing the data using a random process. These models
tend to be maximally disordered and chemically unrealistic.  The information paradigm is closely related 
to the challenge of Materials by Design\cite{eber,emily}, for which one imposes 
external constraints to incorporate additional information on a model to enable
a set of preferred physical properties that are of technological utility.

Neither paradigm is ideal, or even adequate. The simulation paradigm is 
plagued by severe size and time-scale limitations that misrepresent the real 
process of forming a glass, not to mention imperfect interatomic 
interactions.  Despite the development of hardware and software technology 
for distributed-shared-memory computing,  the lack of appropriate 
force-fields or interatomic interactions has been a major obstacle in 
computer simulations of complex multinary glasses. 
For amorphous materials with no or weak {\it glass-forming} ability, either 
approach is rather desperate, and leads to the formation of unrealistic 
models with too many structural defects in the networks. In this paper we introduce
{\it ab initio Force Enhanced Atomic Refinement} (AIFEAR). A preliminary trial of the algorithm using only
empirical potentials recently appeared\cite{fear}.

Others have undertaken related approaches~\cite{pickard, meredig, 
goodwin, epsr, opletal, ecmr, kiran,timilsina}. By including `uniformity' as a constraint 
for the refinement of models, Goodwin and coworkers showed 
their INVERT technique~\cite{goodwin} to produce improved 
models of a-Si and other systems.  A liquid-quench procedure, 
combined with a hybrid Reverse Monte Carlo (HRMC) approach, 
which incorporates both experimental and energy-based constraints has been 
employed by Opletal and coworkers~\cite{opletal}.  
A similar approach via HRMC with empirical bonded and non-bonded forces was used 
by Gereben and Pusztai to study liquid dimethyl trisulfide~\cite{rmc_pot}. 
Likewise, by refining the initial interatomic empirical potential-energy 
function and fitting the input experimental structure-factor data, 
empirical potential structure refinement (EPSR) has been quite successful 
in predicting liquid structures~\cite{epsr}. An alternative approach, 
experimentally constrained molecular relaxations (ECMR), which incorporates 
experimental information in first-principles modeling of materials 
in a `self-consistent' manner\cite{ecmr} was discussed in Ref.\,\cite{ecmr}. 
Recently, a means for including {\it electronic a priori} information 
has also appeared\cite{kiran}. These methods have all contributed 
significantly to the field, yet they have limitations such as employing 
empirical potentials of limited reliability\cite{opletal, fear}, or 
unacceptable convergence properties\cite{ecmr}. {\it A general and 
successful framework for inverting solid state diffraction data does not 
exist. AIFEAR is a major step toward this important goal}.  


We begin with some definitions.  If $V(X_{1} \ldots X_{n})$ 
is the energy functional for atomic coordinates $\{X_i\}$ 
and $\chi^{2}$ measures the discrepancy between diffraction 
experiment and theory, we seek to find a set of atomic 
coordinates $\{X_{i}\}$ with the property that $V$=minimum 
and $\chi^{2}$ is within experimental error. AIFEAR is a very simple
iterative process consisting of (i) producing a structural model at random (at 
an sensible but not necessarily exact density, which may not be available), 
(ii) invoke $N$ accepted 
moves within conventional RMC~\cite{tucker} followed 
by $M$ conjugate-gradient steps using {\it ab initio} 
interactions. We then iterate (ii) until convergence.  The final results do not depend heavily on 
the numerical values of $N$ and $M$, which were chosen 
to be 1000 and 10, respectively, for the present work. 
For the examples discussed here, we find that significantly
fewer {\it ab initio} force calls are needed for AIFEAR than 
for an {\it ab initio} `melt-quench' simulation. In addition, AIFEAR avoids the problem of relative weighting 
of $V$ and $\chi^{2}$ in a penalty or target energy 
functional as in hybrid approaches developed 
elsewhere~\cite{opletal}\cite{sciencepaper}.\footnote[2]{If the 
density of the material is unknown, it is straightforward 
to carry out the simulation at zero pressure (with variable 
cell geometries) in the CG loop, and simply pass the modified 
supercell vectors back to the RMC loop.}

To illustrate the efficacy of this new approach, we begin with 
a persistently vexing problem: the structure of amorphous 
Si~\footnote[3]{Despite its ubiquity in the literature, {\it a}-Si is particularly difficult because the 
network is over-constrained\cite{phillips, thorpe} and it is not a glass former. 
The only methods that yield really satisfactory results are the WWW~\cite{www} 
and ART~\cite{ART} methods. Structural and electronic experiments reveal 
that coordination defects in good quality material have a concentration 
less than a part in 1000. As such, a satisfactory model should have at most 
a few percent (or less) defects. Inversion methods like RMC and {\it ab initio} 
melt-quench both produce unsatisfactory models with far too many coordination 
and strain defects compared to experiments.}. In this illustration, we 
employ the local-orbital-based density functional code {\sc Siesta} 
for the calculation of {\it ab initio} forces, but the approach is 
easily implemented with other {\it ab initio} total-energy codes 
as we show in the next example. 

\begin{figure}
\subfloat[\label{1a}]{%
\includegraphics[width=1.5 in]{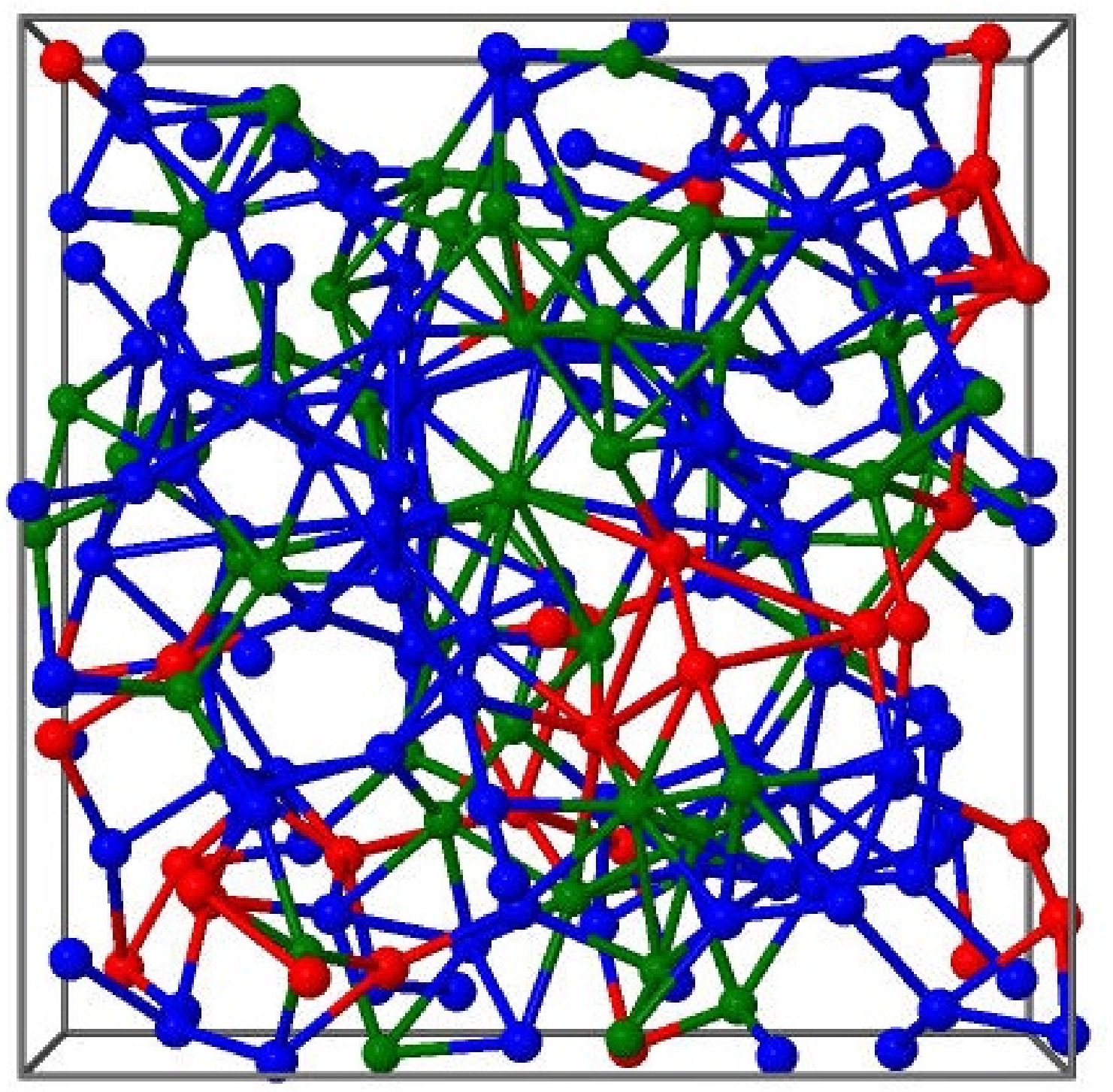}%
 }\hfill
\subfloat[\label{1b}]{%
 \includegraphics[width=1.5 in]{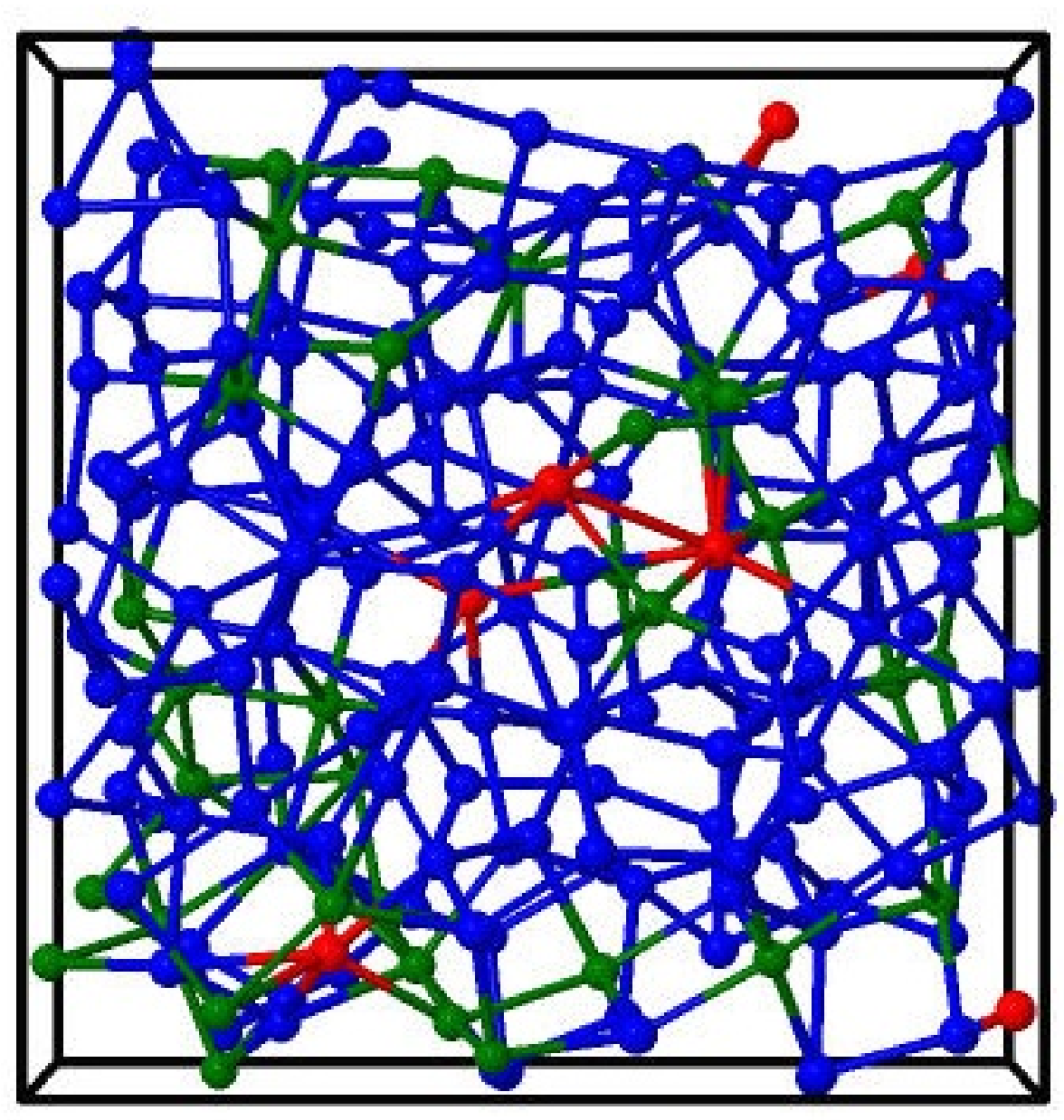}%
}\hfill
\subfloat[\label{1c}]{%
 \includegraphics[width=1.5 in]{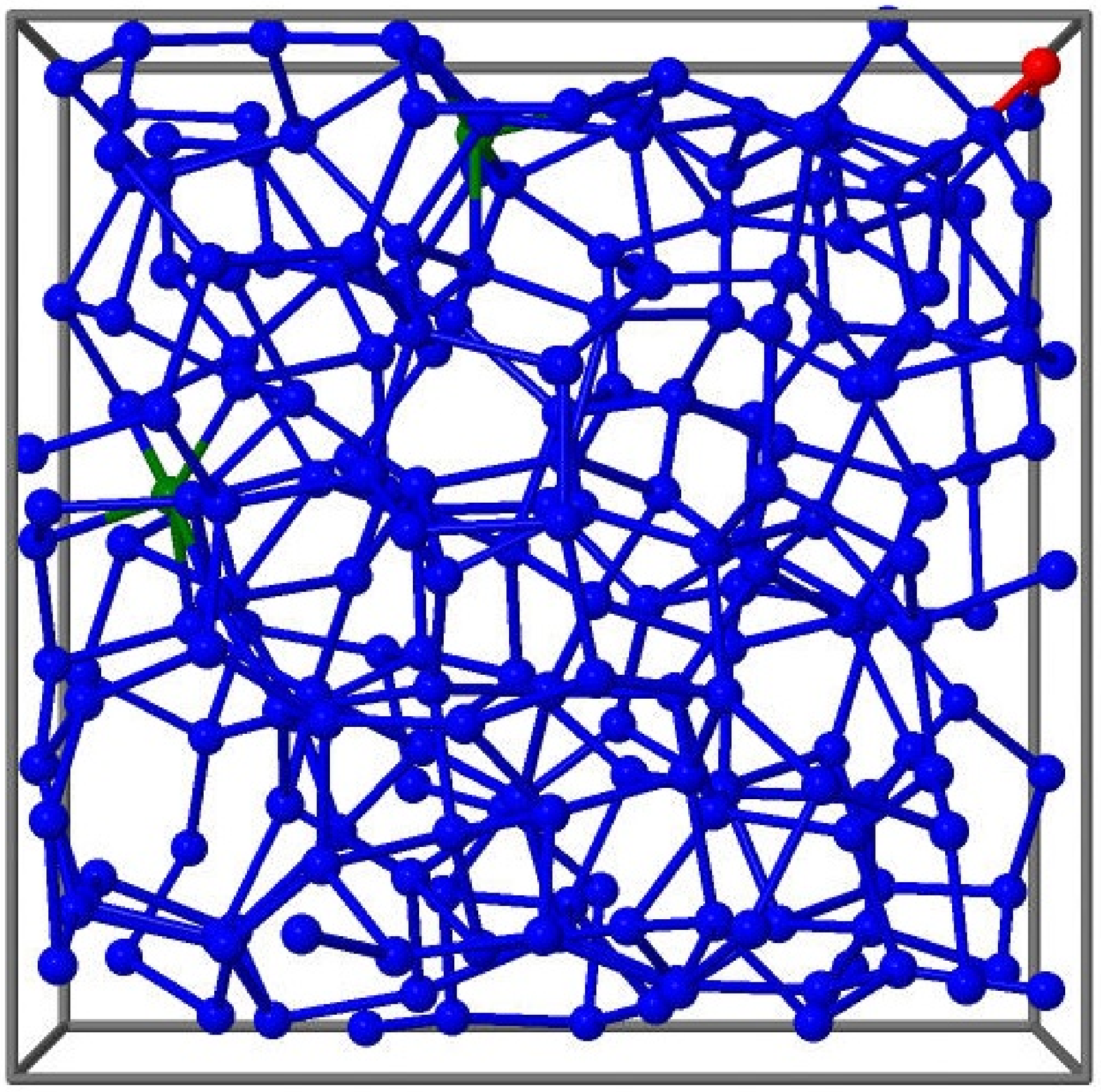}%
 }\hfill
\subfloat[\label{1d}]{%
 \includegraphics[width=1.5 in]{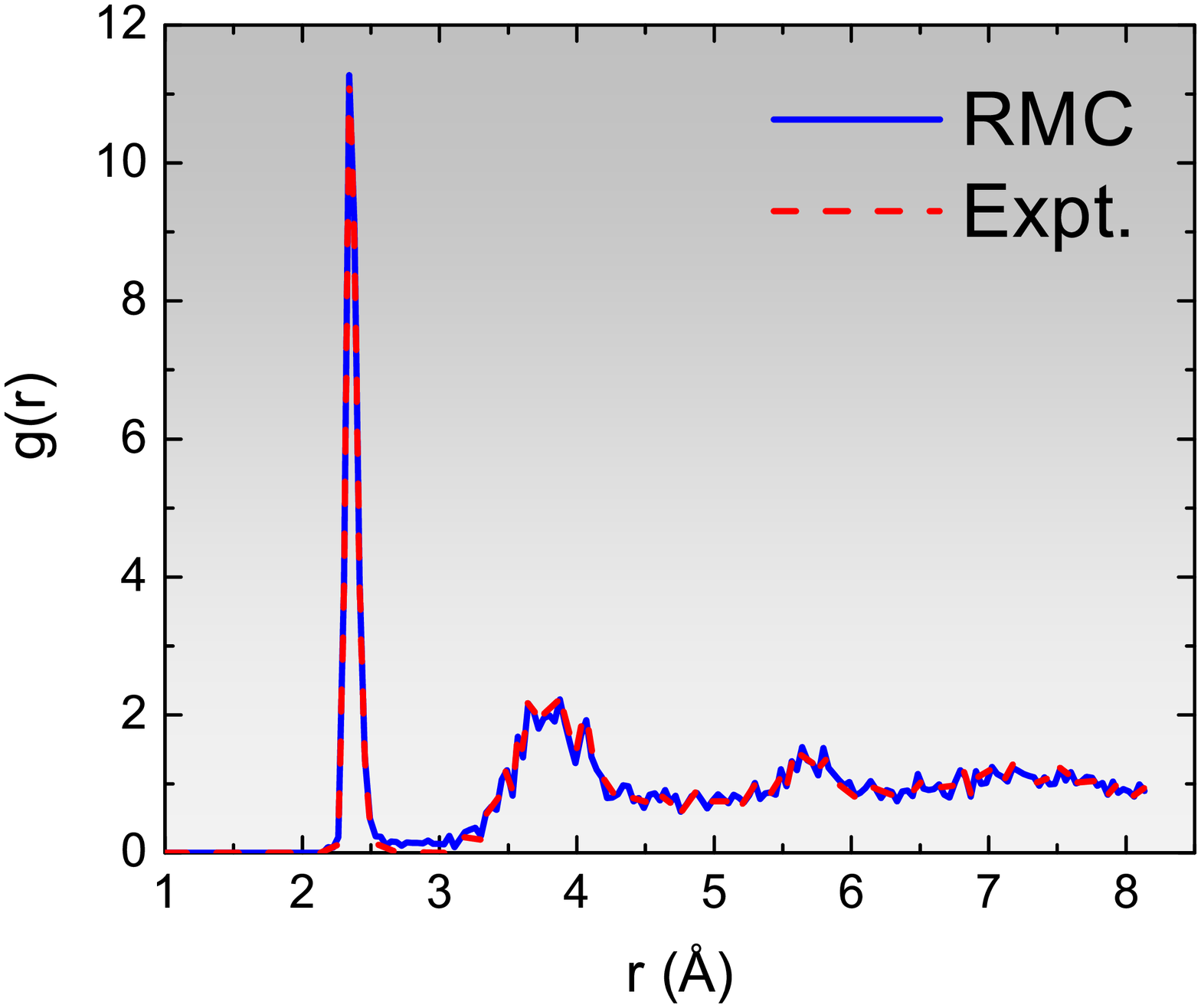}%
 }\hfill
\subfloat[\label{1e}]{%
 \includegraphics[width=1.5 in]{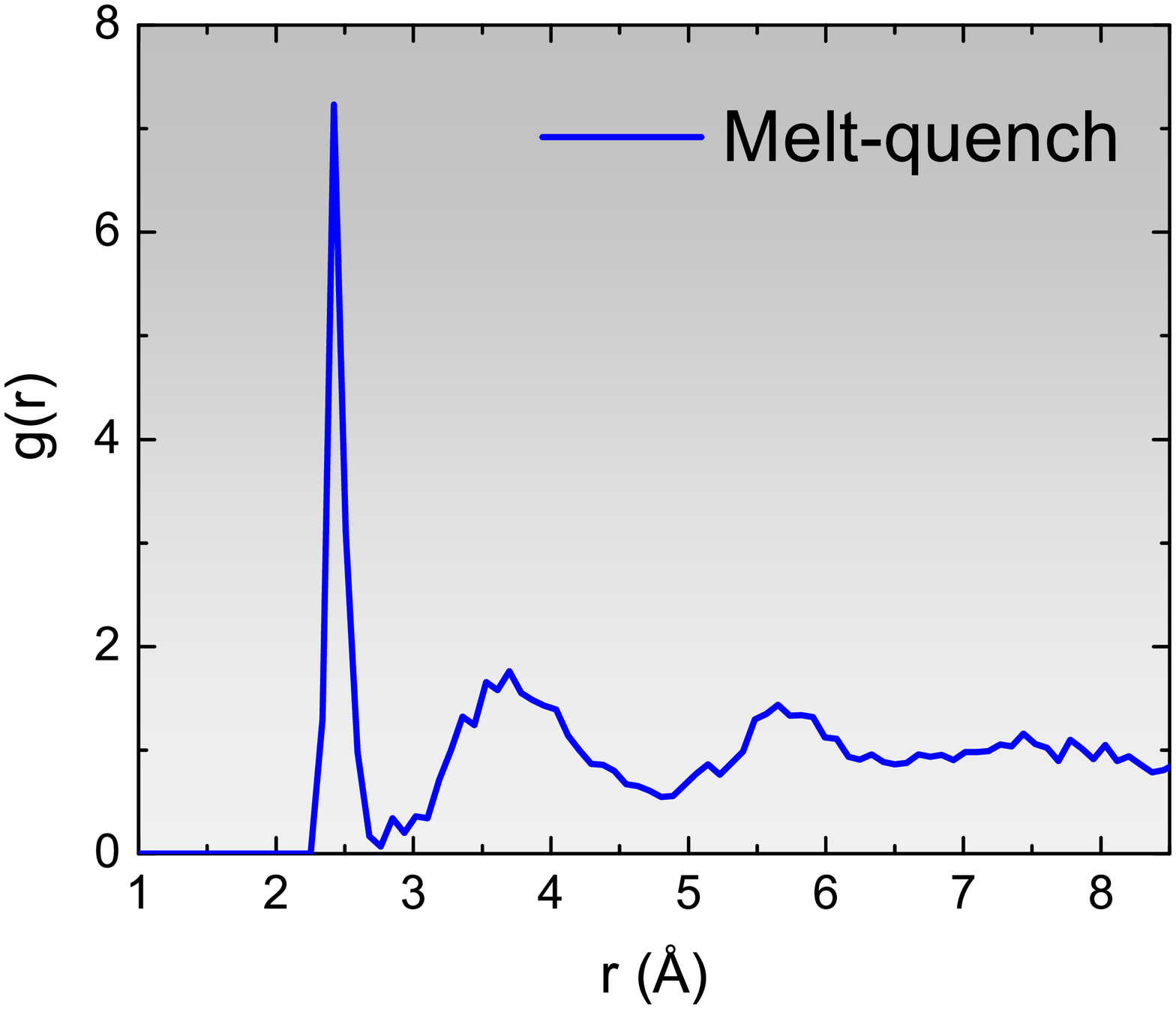}%
 }\hfill
\subfloat[\label{1f}]{%
 \includegraphics[width=1.5 in]{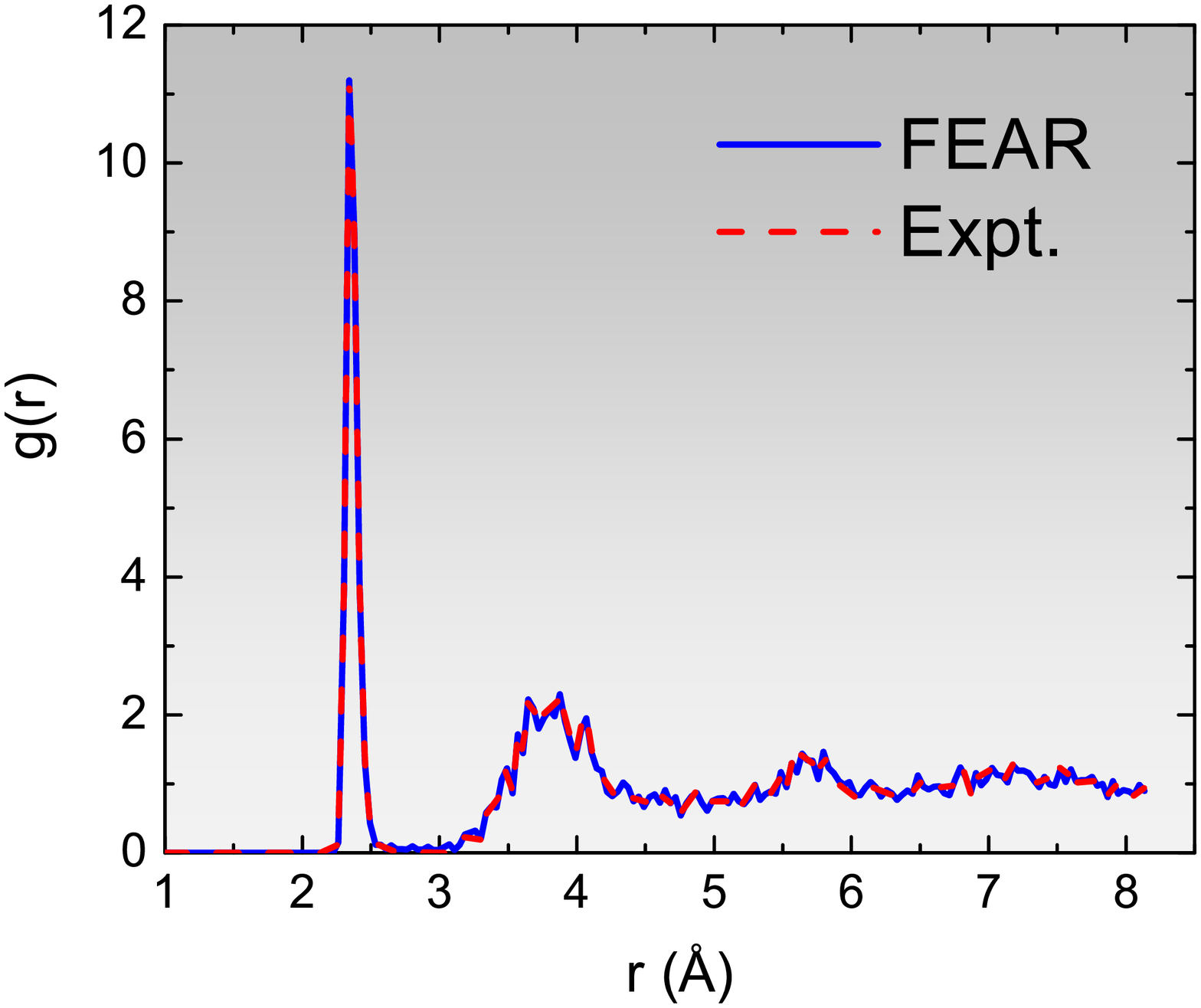}%
 }\hfill
\subfloat[\label{1g}]{%
 \includegraphics[width=1.5 in]{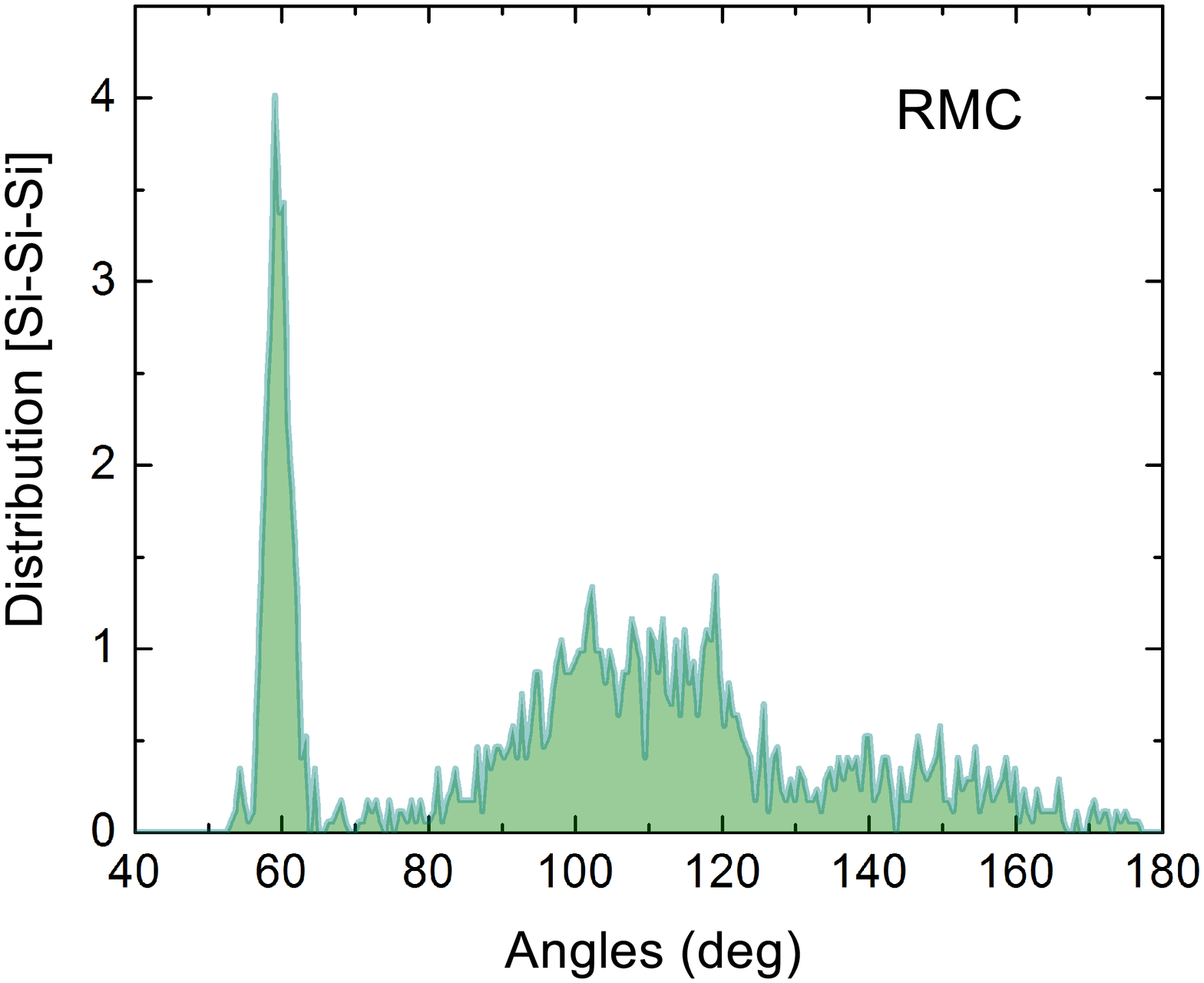}%
 }\hfill
\subfloat[\label{1h}]{%
 \includegraphics[width=1.5 in]{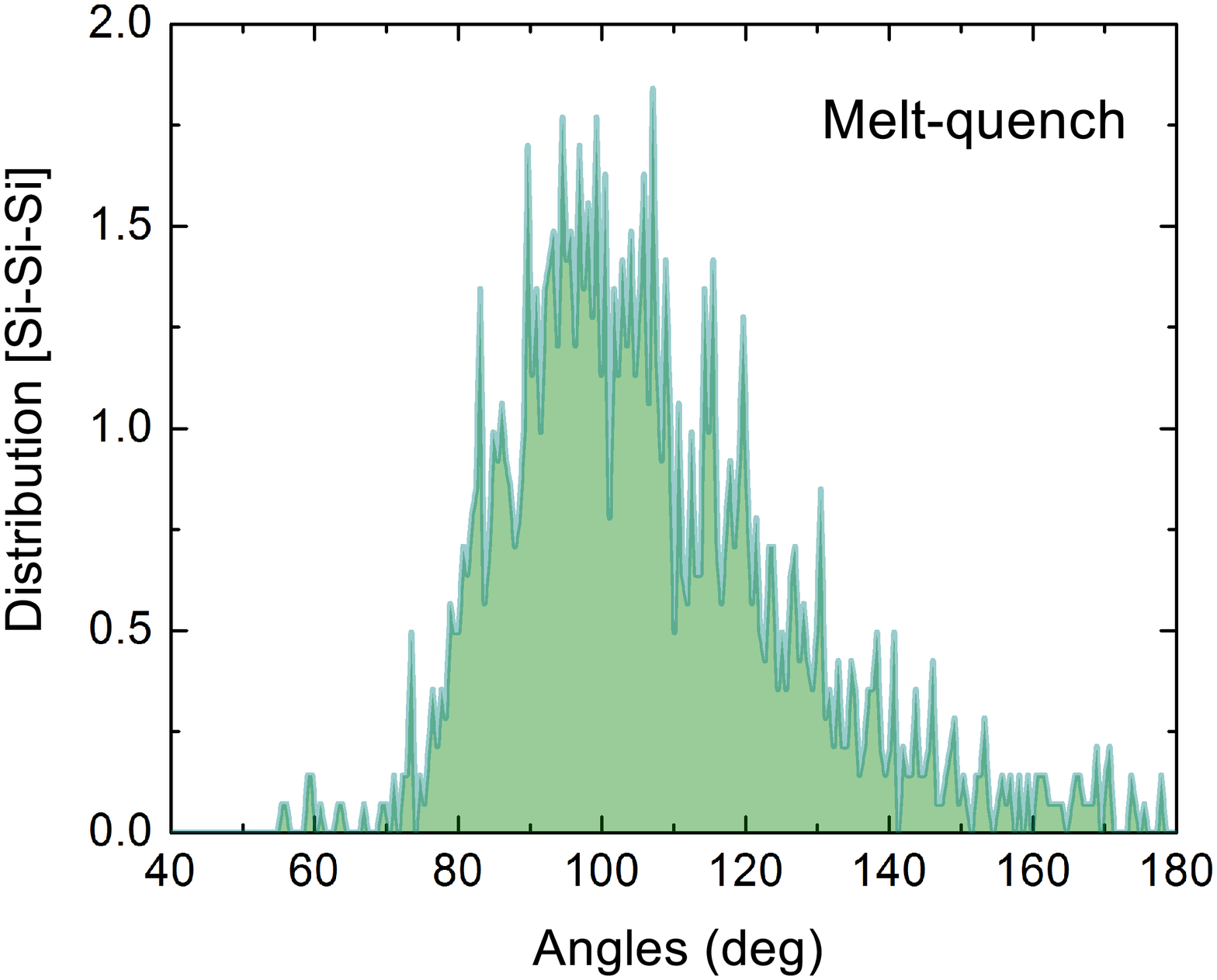}%
 }\hfill
\subfloat[\label{1i}]{%
 \includegraphics[width=1.5 in]{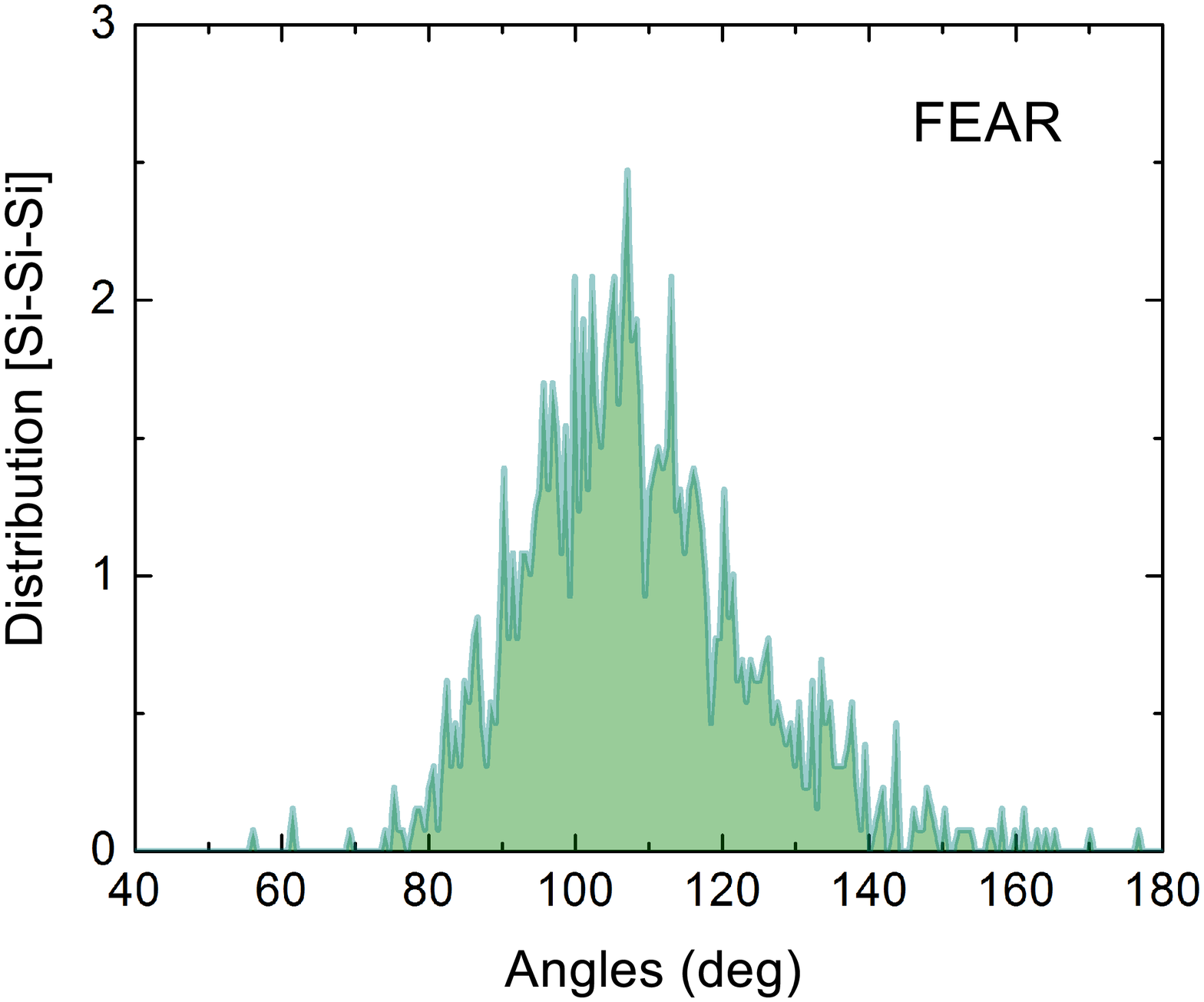}%
 }
\caption{Top: A 216-atom model of {\asi} obtained from (a) RMC, (b) 
melt-quench and (c) {\it ab initio} FEAR simulations. Silicon 
atoms with a coordination number of 3, 4 and 5 are shown in 
red, green, and blue colors, respectively. 
Center: The radial distribution function (RDF) for the 
(d) RMC, (e) melt-quench and (f) {\it ab initio} FEAR models.
Bottom: The bond-angle distributions for the models as indicated 
in the figure. See supplementary materials for animations 
showing the formation of three-dimensional network structure 
and the corresponding evolution of the radial and 
coordination-number distributions. 
}

\label{asi1}
\end{figure}

\begin{figure}
\begin{center}
\subfloat[\label{3a}]{%
\includegraphics[width=3.2 in]{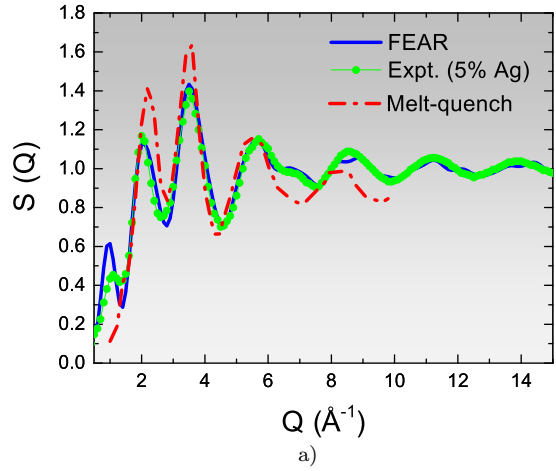}%
 }\hfill
\subfloat[\label{3b}]{%
 \includegraphics[width=3.2 in]{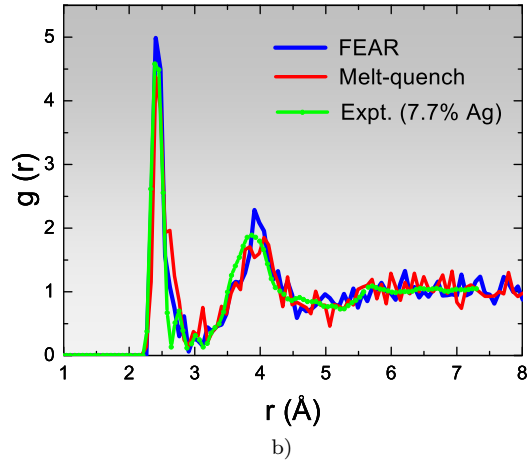}
 }
\caption{
(a) Structure factors of (GeSe$_3$)$_{1-x}$Ag$_{x}$ [$x$=0.05] from 
{\it ab initio} FEAR. Experimental data, from neutron diffraction 
measurements, are shown for comparison~\cite{pradel}. Melt-quench 
data are from Pradel {\it et al.}\cite{pradel}
(b) The radial distribution function of (GeSe$_3$)$_{1-x}$Ag$_{x}$ [$x$=0.077]
from {\it ab initio} FEAR and melt-quench simulations. Experimental RDF shown 
here are from Zeidler {\it et al.}~\cite{anita}. 
}

\label{geseconv}
\end{center}
\end{figure}

We began by preparing three 216-atom models of {\it a}-Si, 
at the experimental density of 2.33 g.cm$^{-3}$ \cite{mcguire}, 
using (1) RMC, (2) melt-quench, and (3) {\it ab initio} FEAR.  
The starting atomic configurations were chosen to be {\it random}, and the 
diffraction data from Ref.\,~\cite{thorpe} were employed in RMC and 
{\it ab initio} FEAR.  The structural properties of {\it a}-Si, obtained from these 
models, are summarized in Fig.\,\ref{asi1}. For a discussion on 
convergence and comparisons to other calculations, see Methods 
section (cf. Fig.\,\ref{wwwfig}). 
RMC produces a highly unrealistic model, far from the accepted tetrahedral 
network topology, as seen in Fig.\ref{asi1}. Melt-quench, while better, still produces far too 
many coordination defects. By contrast, AIFEAR produces a nearly perfect structure, with 99.07\% fourfold coordination, 
and a bond-angle distribution close to that of from a 
WWW model.\footnote[4]{ 
In comparing the bond-angle distributions (from AIFEAR with that of 
from WWW), one must take into account the fact that {\it ab initio} 
interactions tend to produce a slightly wider bond-angle distribution.
This partly explains the large value of the RMS deviation observed 
in our works. 
} 
We wish to emphasize that the starting configuration 
used in AIFEAR was {\it random}, so that one can logically infer that 
a combination of atomic-radial-correlation data and DFT interactions 
leads to an almost perfect tetrahedral network as illustrated in 
Fig.\,\ref{asi1}. Table 1 lists the key structural properties of 
the model, along with the total energy per atom. 
In the Methods section, we report 
the detailed convergence of total energy $E$ and $\chi^{2}$. 
In the Supplementary Materials, we also offer an animation of the 
convergence of {\it ab initio} 
FEAR by showing the formation of a near perfect tetrahedral network 
as the simulation proceeds with the disappearance of coordination 
defects 
\begin{table}
\caption{ 
Total energy and key structural properties of {\it a}-Si models. 
The energy per atom is expressed with reference to the 
energy of the WWW model. 
}
 \begin{center}
\begin{tabular}[b]{|p {1.8cm}|p {1.4cm}|p {1.4cm}|p {1.4cm}|p {1.2cm}|}

  \hline
    &RMC & Melt-quench & AIFEAR & WWW\\ \hline
    4-fold Si (\%) & 27 & 80 & 99.07 &100 \\ \hline
     SIESTA energy (eV/atom) & 3.84  &0.08 & 0.03 &0.00\\ \hline
     Average bond angle (RMS deviation)  & 101.57{\deg} (31.12\deg)  & 107.04{\deg} (20.16\deg)& 108.80{\deg} (14.55\deg) &108.97{\deg} (11.93\deg)\\ \hline
      \hline
\end{tabular}
\label{tab1}
\end{center}
\end{table}
For a challenging and timely example, we have also studied the solid 
electrolyte material {\aggese3}. This is a chemically complex system 
with important applications to conducting bridge computer FLASH memory 
devices, which are of considerable fundamental and technological interest. 
We employ the same scheme as for {\asi}, but with {\it ab initio} 
interactions from the plane-wave DFT code VASP\cite{kresse1,kresse2,kresse3}, with 135 and 108 atoms 
in a unit cell of length 15.923 {\AA} and 15.230 {\AA} for $x=0.05$ and $x=0.077$, 
respectively. These values correspond to the densities of 4.38 g.cm$^{-3}$ 
and 4.04 g.cm$^{-3}$ for the models with 5\% and 7.7\% Ag, respectively. For $x=0.05$, 
both the structure-factor data and density of 4.38 g.cm$^{-3}$ are taken 
from the work of Piarristeguy and Pradel~\cite{pradel}. For $x=0.07$, we 
have used the RDF data provided by Zeidler and Salmon\cite{anita}, 
and a density of 4.04 g.cm$^{-3}$ was obtained from a zero-pressure 
conjugate gradient relaxation using {\sc Vasp}. For completeness, we have also studied a 
melt-quench model of $x=0.077$ as described in the Methods 
section. The melt-quench model (in Fig.\,\ref{geseconv}a) shows significant discrepancies with 
experiments: the first sharp diffraction peak (FSDP) near 1 {\AA$^{-1}$} is 
absent, and there are significant inconsistencies in the structure factor 
at high $k$ values~\footnote[5]{The FSDP is an indicator of medium range order, 
a signature of structural correlations between the tetrahedral GeSe structural 
building blocks of the glass.}. By contrast, the AIFEAR model captures all 
the basic characteristics of the structure factor, including the FSDP (in fact, it 
slightly {\it overfits} the FSDP). We show that the method has similar 
utility in either real or $k$ space, using $S(k)$ for the first composition 
and $g(r)$ for the second.  Figure \ref{geseconv} shows the structure factors
and radial distribution functions obtained from AIFEAR and melt-quench simulations, 
and compares with the experimental data from neutron diffraction 
measurements~\cite{pradel, anita}. 

The GeSeAg systems are of basic interest as solid electrolytes. 
One of the most interesting questions pertains to the dynamics 
of Ag atoms, which are sufficiently rapid that they can be tracked 
even in first-principles molecular-dynamics simulations~\cite{dad_gese3ag}. The fast Ag dynamics have led to 
the invention of conducting bridge Random Access Memory (CBRAM)~\cite{patent,waser}. 
As this dynamics appears to be of trap-release form~\cite{dad_gese3ag}, the 
structure, including features like medium range order, and associated 
energetics may be expected to play a key role in the silver hopping. The 
7.7\% Ag composition is near to a remarkable and abrupt ionic mobility 
transition~\cite{transitions1,transitions2}. Dynamical simulations are currently 
underway to determine the role of the structure in this dynamics.

The following features of {\aggese3} glasses have been observed in the AIFEAR model:  
1) The Ge-Se correlation is not affected by an increase in Ag content: Ge(Se$_{1/2}$)$_{4}$ 
tetrahedra remain the fundamental structural units in the network.  2)  
Ge-Ge correlations, greatly affected by Ag doping, are revealed 
by the shift in Ge-Ge nearest-neighbor distance from 3.81 {\AA} in 
Ag=0\% \cite{pradel} to 2.64 {\AA} and 2.56 {\AA} in 
Ag=5\% and Ag=7.7\% respectively, supporting the argument of Ag being the 
network modifier.  3) The Ag-Se correlation peak is near 
2.66 {\AA} for both the systems, which is consistent with the 
experimental work of Zeidler~\cite{anita} and others~\cite{pradel}.
4) The Se-Se coordination number for 5\%  and 7.7\% Ag are 
1.12 and 0.83 (0.81 from experimental data~\cite{anita}), respectively. 
This is consistent with the observed phenomena of decrease in 
Se coordination with the increase in Ag concentration~\cite{pradel}.

\begin{figure}
\subfloat[\label{5a}]{%
\includegraphics[width=2.5 in]{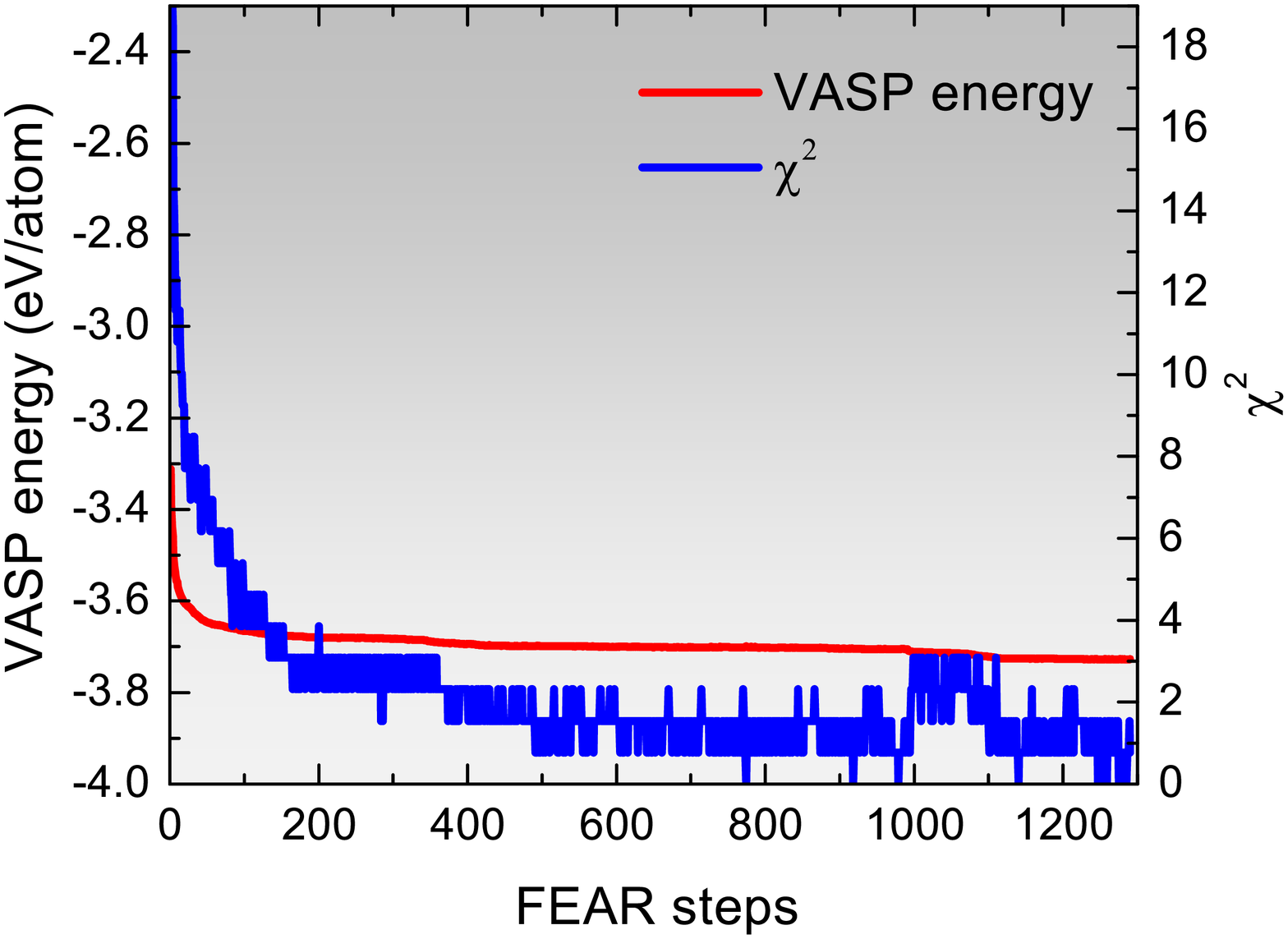}%
 }\hfill
\subfloat[\label{5b}]{%
\includegraphics[width=2.5 in]{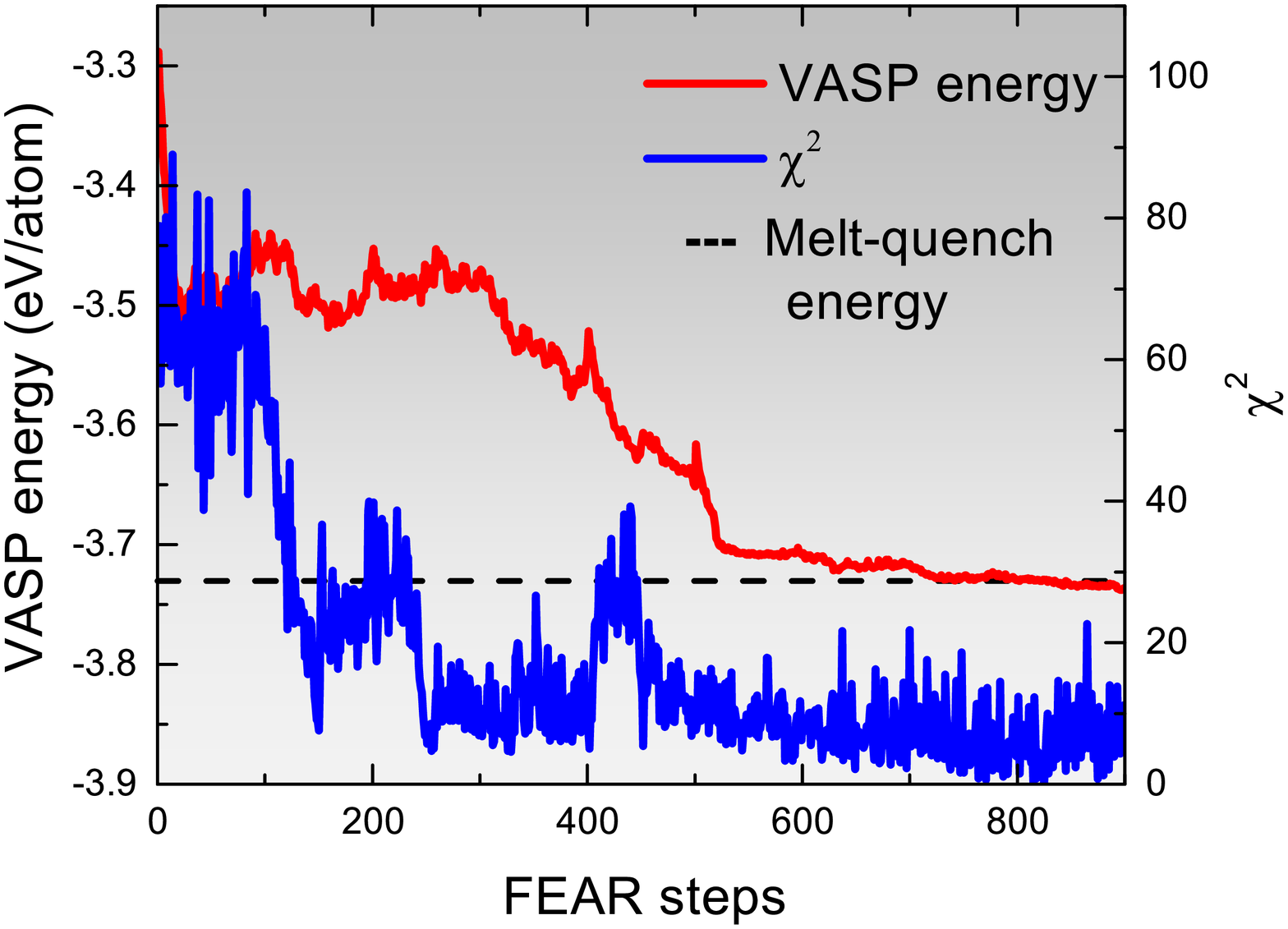}
}
\caption{Total energy per atom and the cost function ($\chi^{2}$) 
versus AIFEAR steps for two models with (a) 5\% and (b) 7.7\% Ag-doped 
GeSe$_{3}$. The melt-quench energy for the 7.7\% Ag model is indicated 
for comparison.}
\label{fig5}
\end{figure}

\begin{figure}
\begin{center}

\subfloat[\label{2a}]{%
\includegraphics[width=2.2 in]{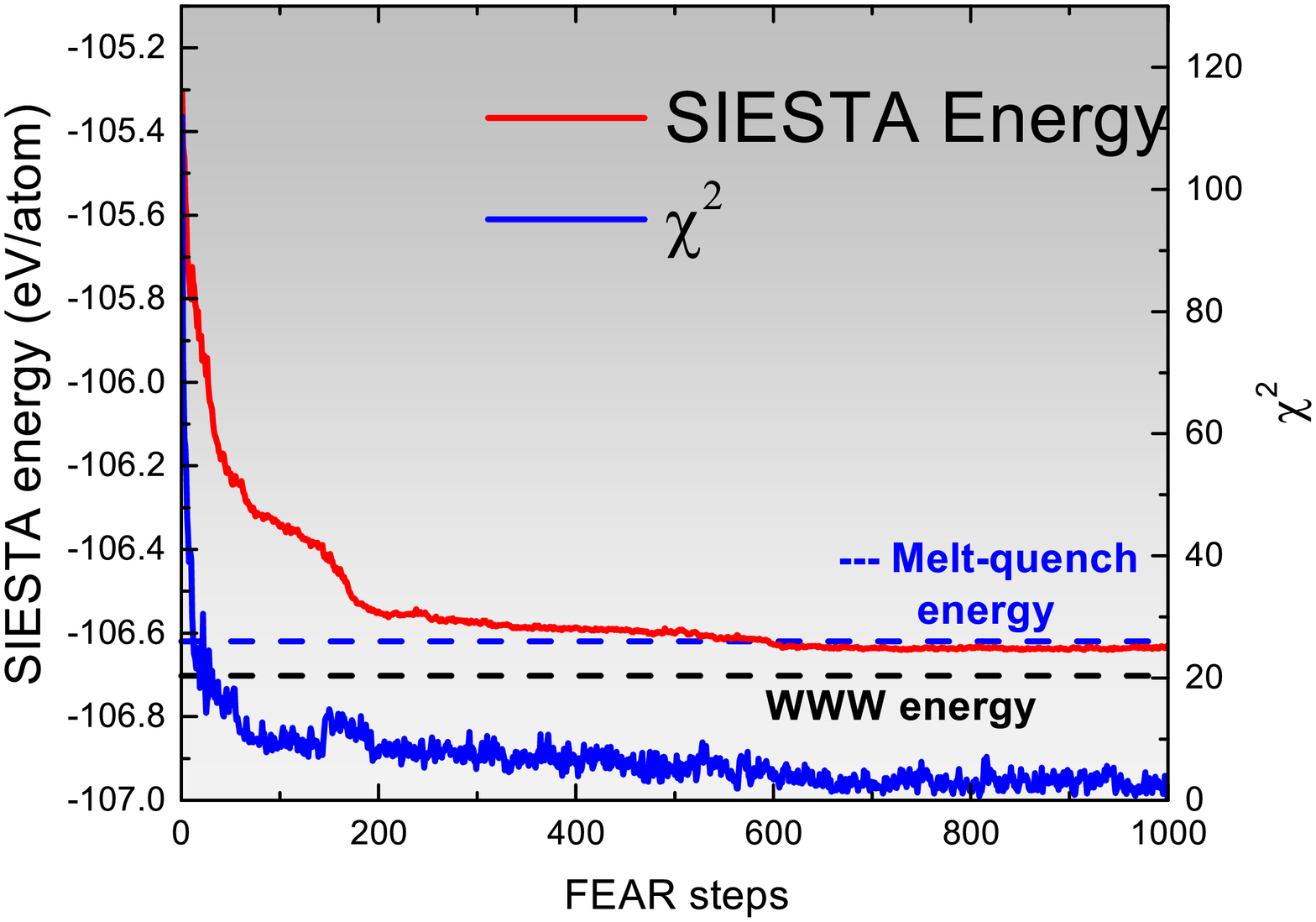}%
 }\hfill
\subfloat[\label{2b}]{%
\includegraphics[width=2.2 in]{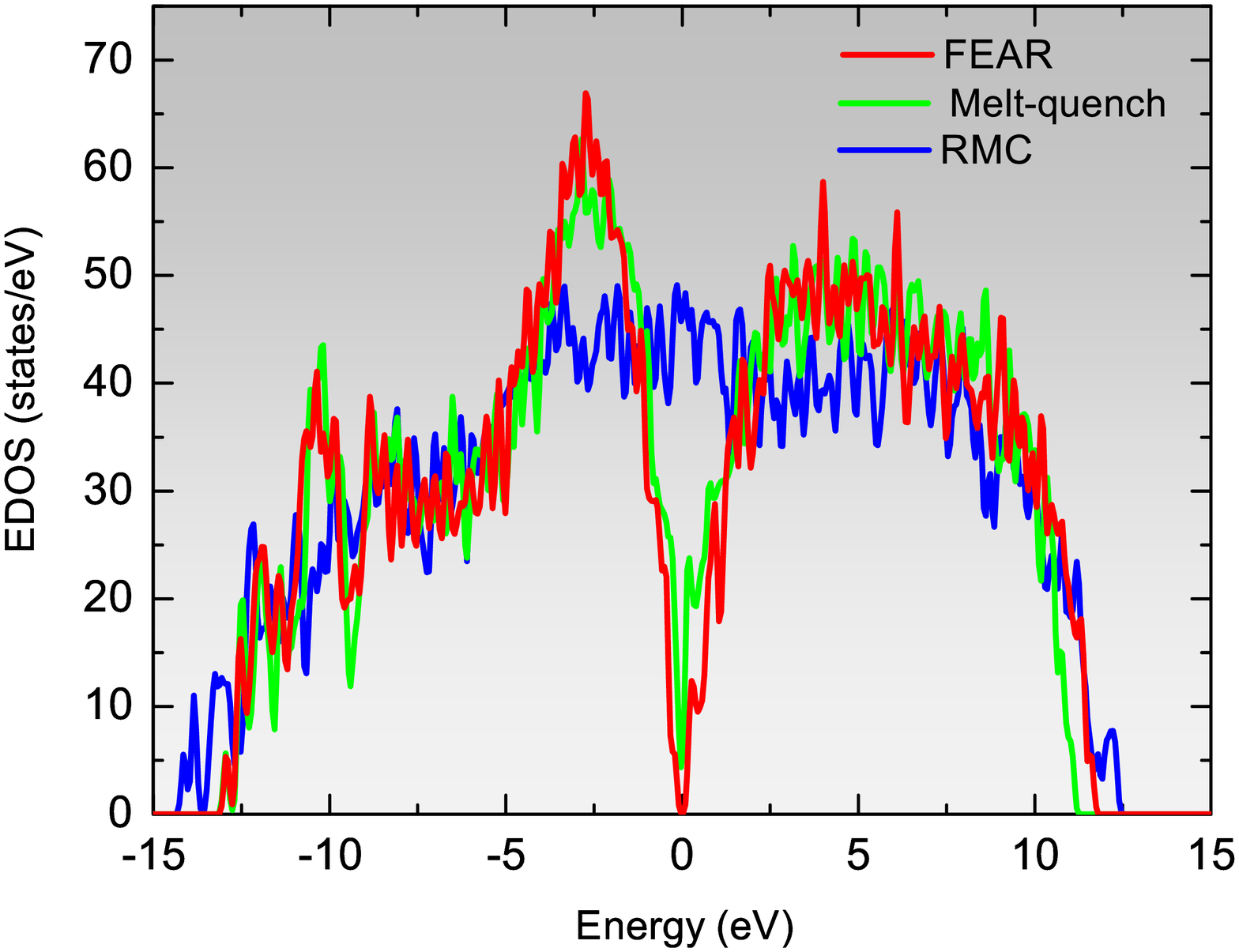}%
 }\hfill
\subfloat[\label{2c}]{%
 \includegraphics[width=2.2 in]{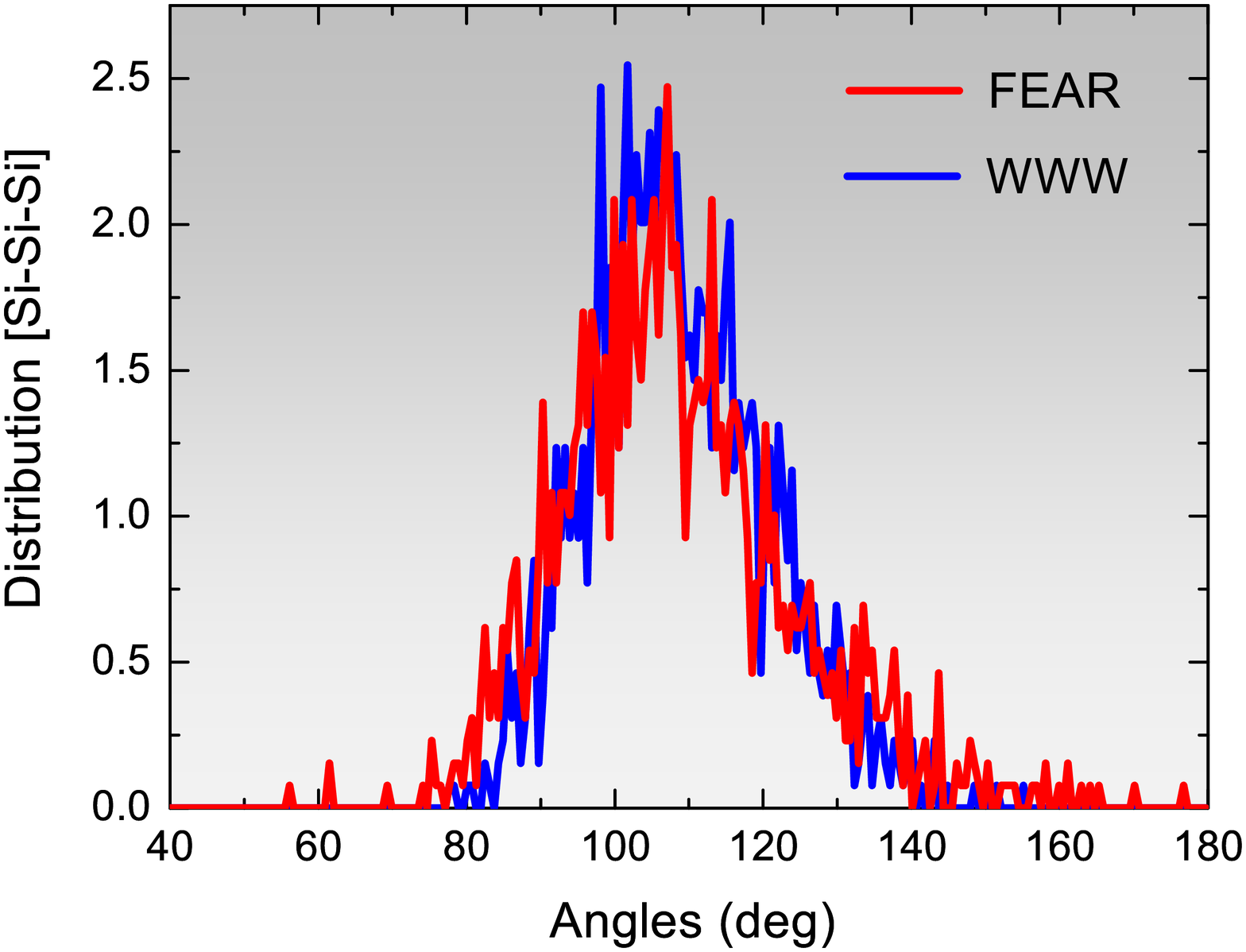}
 }
\caption{
Results for 216-atom {\asi}: (a) The variation of cost function and total energy with the number 
of AIFEAR steps. 
(b) Electronic density of states (EDOS) for RMC, melt-quench 
and AIFEAR models with the Fermi level at 0 eV. (c) The bond-angle distribution 
from AIFEAR compared to that of WWW (see Table 1 for details). 
}

\label{wwwfig}
\end{center}
\end{figure}

Beside retaining the important chemical features of the network, 
the AIFEAR model is superior to the melt-quench model by the manifestation of
a prominent FSDP (cf. Fig.\,\ref{3a}),  a signature of medium 
range order in these materials. Absence of the FSDP 
indicates the lack of structural correlations in the Ge(Se$_{1/2}$)$_{4}$ 
tetrahedra, which is less prominent for low Ag concentration. Also, 
the energy of the AIFEAR model for x=0.077 is 0.02 eV/atom less than 
the melt-quench model (see Fig.\,\ref{5b}). 

It is important and promising that in the GeSeAg systems, as in {\it a}-Si, 
AIFEAR is not a greedy optimization scheme, as it is evidently able to unstick 
itself (for example in Fig.\ref{5b}) near 400 steps, there is a dramatic 
and temporary increase in $\chi^{2}$, which then enabled the system to find 
a new topology which enabled further reduction of both $\chi^{2}$ and $E$. A 
similar, if less dramatic, event is indicated in Fig.\ref{5a} around step 
1100). The Monte-Carlo moves robustly explore the configuration space and 
are not so prone to getting trapped as MD simulations, and yet 
the chemistry is properly included in the {\it ab initio} relaxation loop.

In conclusion, we have introduced a new and practical method that enables 
the joint exploitation of experimental information and the information 
inherent to {\it ab initio} total-energy calculations, and a powerful new 
approach, to the century-old problem of structural inversion of diffraction data. 
The method is simple and robust, and independent of the systems, the 
convergence of which has been readily obtained in two highly distinct 
systems, both known to be challenging and technologically useful. By 
direct calculation, we show the network 
topology implied by pair correlations and accurate total energies: an 
essentially fourfold tetrahedral network, structurally similar to WWW 
models, including the bond-angle distribution. Using only the total 
structure factor (or pair-correlation) data and {\sc Siesta}/{\sc Vasp}, 
we obtain models of unprecedented accuracy for a difficult test case ({\it a}-Si) 
and 
a technologically important memory material (GeSe$_3$Ag). 
The inclusion of {\it a priori} experimental information emphasized here may also be 
developed into a scheme to include other information for materials 
optimization. It is easily utilized with any interatomic potentials, 
including promising current developments in ``machine learning"~\cite{nat_mach_learn}. 
The method is unbiased in the sense that it starts from a completely 
random configuration and explore the configuration space of a total-energy 
functional aided by additional experimental information to arrive 
at a stable amorphous state. Beside these attributes, it requires fewer force calls to the 
expensive {\it ab initio} interactions.

{\bf Acknowledgements}
We thank the US NSF under the grants DMR 150683, 1507166 and 1507670 for supporting 
this work, and the Ohio Supercomputer Center for computer time. We 
thank Dr. Anita Zeidler and coworkers providing us with their experimental 
data.

{\bf Methods}
As described in the main text, AIFEAR jointly minimizes the 
configurational energy $V$ and\cite{mcgreevy,rmc2} %
\be
\chi^2 = \sum\limits_{i}\left[\frac{F_E(k_i)-F_M(k_i)}{\sigma(k_i)}\right]^2, 
\label{eqn1}
\ee
where $F_{E/M}(k_i)$ is the experimental/model structure factor, 
and $\sigma(k_i)$ is the error associated with the experimental 
data for wave vector $k_i$.  To undertake this program, (i) we 
begin with a random model, (ii) invoke $M$ RMC accepted moves 
followed by $N$ conjugate-gradient steps to optimize the total energy. We have 
found $M=1000$ and $N=10$ to be satisfactory for the materials of this paper.  The process (ii) is repeated 
until the desired accuracy of $\delta\chi^{2}\approx$ 0.1 and, 
a force tolerance of $\delta f\approx$ 0.02 eV/{\AA} is attained. 
All that is required are RMC and total-energy codes and an 
appropriate driver program connecting them.

{\bf Amorphous Si} Initially, conventional RMC (i.e. without any constraint) 
was performed using the RMCProfile software~\cite{tucker} for a {\it random} 
starting configuration of 216-atom {\asi} with a cubic box of side 16.281 {\AA} corresponding to the density 2.33 
g.cm$^{-3}$. 
The maximum step length of the RMC moves for Si atoms is chosen to be 
0.05 {\AA}.  In a parallel simulation, the same starting configuration 
is taken through a process of melt-quench using the density-functional 
code {\sc Siesta}~\cite{siesta} with single-$\zeta$ basis under Harris 
functional scheme~\cite{siesta} within the local density approximation. 
The total-energy and force calculations are restricted to the $\Gamma$ 
point of the supercell Brillouin zone. After melting at 2300 K, the 
liquid structure was quenched to 300 K at a rate of 240 K/ps. Each step was 
followed by the equilibration of the system for 2000 time steps. Finally, 
the configuration is subjected to {\it ab initio} FEAR simulations 
with the same Hamiltonian and ``data". To ensure the reproducibility of 
the method, we have modeled 10 {\asi} models starting from random 
configurations and the models yielded 4-fold coordination exceeding 
96\%. Details of convergence and comparison to the best available WWW 
model is provided in Fig.\,\ref{wwwfig}. The elimination of defects 
is chronicled in an animation provided in the Supplementary Materials. 

{\bf Chalcogenide glasses} The experimental structure factors data 
taken from the work of Piarristeguy {\em et al.}\cite{pradel} for 5\% Ag 
and the pair distribution function (PDF) was obtained from the work 
of Zeidler and Salmon~\cite{anita} for 7.7\% Ag are incorporated in 
the plane-wave basis DFT code {\sc Vasp}~\cite{kresse1,kresse2,kresse3} 
using projected augmented plane waves (PAW)~\cite{paw} with Perdew-Burke-Ernzerhof (PBE) 
exchange-correlation functional~\cite{pbe} and a plane-wave cutoff 
of 312.3 eV. All calculations were carried out at $\Gamma$ point. 
The {\em random} starting configurations of 5\% and 7.7\% Ag-doped {\gese3ag} 
were subjected to {\abfear}. The 5\% Ag-doped {\gese3ag} {\abfear} models are 
compared to the melt-quench model of the identical system of Piarristeguy 
and co-workers~\cite{pradel}. The melt-quench model of 7.7\% Ag-doped 
GeSe$_{3}$ model is prepared by melting the same starting configuration 
at 1400 K for 10,000 steps followed by a quenching to 300 K at the rate 
of 100 K/ps, and then by equilibrating at 300 K for another 5000 steps. 
To estimate the density the equilibrated system, the volume of the simulation 
cell was relaxed.  A final relaxation at zero pressure was employed, which 
yielded a density of 4.04 g.cm$^{-3}$. Throughout the calculations, we have 
used a time step of 1.5 ps. 

We have included a comparison of the number of force calls in the various 
simulations in Fig.\,\ref{fig5}. It is evident from Fig.\,\ref{fig5} that 
AIFEAR offers a significant computational advantage, with fewer force 
calls to the expensive {\it ab initio} codes. 

%
\begin{figure}
\begin{center}
\includegraphics[width=3.0 in]{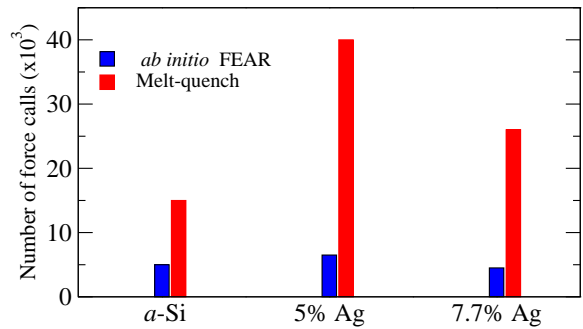}
\hfill
\caption{
Comparison of number of force calls in {\abfear} 
with melt-quench simulations for {\asi}, and 
5\% and 7.7\% Ag-doped GeSe$_{3}$. Note that 
the number of force calls in melt-quench 
simulations vary considerably for different 
systems.
}
\label{timefig}
\end{center}
\end{figure}
{\bf{Supplementary Materials}}
Animations are provided revealing the detailed process of formation of tetrahedral {\it a}-Si 
networks from random to converged in {\it ab initio} FEAR simulations. These include 
the AIFEAR radial distribution function and coordination-number distribution. 


\end{document}